\documentclass[12pt]{article}

\usepackage{mypackage}
\usepackage{amsmath}
\usepackage{footmisc}
\usepackage{comment}
\usepackage{dsfont}
\usepackage{color}
\usepackage{pdfpages}
\usepackage{graphicx}
\usepackage{amssymb}
\usepackage{amsfonts}
\usepackage{hyperref}
\usepackage{tikzsymbols}
\usepackage{latexsym,amsthm,amsmath,amssymb,url}
\usepackage{mathrsfs}
\usepackage{amsfonts}
\usepackage{graphics}
\usepackage[round]{natbib}
\usepackage{enumitem}
\usepackage{threeparttable}
\usepackage{float}
\usepackage{adjustbox}
\usepackage{pdflscape}
\usepackage{pdfpages}
\hypersetup{
    linkcolor = blue
}

\usepackage{amsmath,amssymb,amsthm,array,booktabs,bookmark,bm,color,enumerate,graphicx,multirow,subfigure}

\usepackage{natbib}
\usepackage{har2nat}
\setcitestyle{authoryear} %Citation-related commands

\usepackage{geometry}
\usepackage{changepage}

\begin{document}
\date{\today}

\title{Can the decoy effect increase cooperation in networks? An experiment\thanks{%
CC acknowledges funds from City St George's University of London (Pump-priming funding). PP acknowledges PRIN grants P20228SXNF
and 2022389MRW financed by the Italian Ministry of Research. We thank Tommaso Batistoni for running the experimental sessions. }}
\author{Claudia Cerrone\protect\thanks{City St George's, University of London, Email: \href{mailto:claudia.cerrone@city.ac.uk}{claudia.cerrone@city.ac.uk}} \and Francesco Feri\protect\thanks{Royal Holloway, University of London, Email: \href{mailto:francesco.feri@rhul.ac.uk}{francesco.feri@rhul.ac.uk}} \and Anita Gantner\protect\thanks{University of Innsbruck, Email: \href{mailto:anita.gantner@uibk.ac.at}{anita.gantner@uibk.ac.at}} \and Paolo Pin\protect\thanks{Università di Siena \& BIDSA, Università Bocconi, Email: \href{mailto:paolo.pin@unisi.it}{paolo.pin@unisi.it}}}

\maketitle

\begin{abstract}
This paper investigates whether the decoy effect -- specifically the attraction effect -- can foster cooperation in social networks. In a lab experiment, we show that introducing a dominated option increases the selection of the target choice, especially in early decisions. The effect is stronger in individual settings but persists in networks despite free-riding incentives, with variation depending on the decision-maker's strategic position.\end{abstract}

\section{Introduction}

The decoy effect is a cognitive bias where introducing a clearly inferior option increases the likelihood of choosing a specific original option, violating the axiom of regularity \citep{Luce1997}. Empirical work shows that this bias can systematically shift preferences \citep{Huber1982, Simonson1989}. The two main variants of the decoy effects are the attraction effect, where an asymmetrically dominated decoy increases the appeal of the target option, and the compromise effect, where the decoy makes the target seem a balanced choice or middle ground.

While well-documented in individual, non-strategic contexts like marketing, the decoy effect is less understood in interactive environments. A few studies suggest it may aid coordination \citep{Colman2007, Amaldoss2008} or influence social dilemma choices \citep{Wang2018}, with recent work studying the decoy effect in bargaining settings \citep{Galeotti2019}.

In this paper, we contribute to this emerging literature by studying whether the decoy effect, and in particular the attraction effect may foster cooperation in strategic network settings. Consider a scenario during a virus outbreak: individuals can protect themselves and others by getting vaccinated or wearing a face mask, both of which have positive spillovers. If compliance is low, introducing an inferior alternative—like a vaccine with more side effects or a face shield that is less comfortable—can make the effective option (e.g., mask-wearing) more appealing. According to the attraction effect, this increases uptake of the protective behaviour.

This paper investigates whether introducing a decoy option can promote cooperation and socially beneficial choices in networks. In our experiment, participants are placed in fixed 6-member networks where the risk of a monetary loss increases with the number of neighbours. Buying protection reduces this risk and may also benefit others.
We vary two factors: (1) the presence of a decoy option -- an inferior protection that is more expensive and less effective than the target -- and (2) the presence of network externalities. In the network treatments, one's choice affects both own and neighbours’ risk, creating free-riding incentives. In the individual treatments, choices only affect personal risk.
Each treatment includes 10 rounds to observe decision patterns over time.

We find that in the first round, before any learning occurs, the presence of a decoy significantly increases the likelihood of choosing the target option, both in individual and network settings. Over all rounds, the decoy no longer affects target choice specifically, but it does increase the overall likelihood of purchasing some form of protection. The effect on the target choice remains significant in the individual setting.

The decoy effect is stronger in individual environments, likely due to the absence of free-riding. In networks, strategic considerations and externalities weaken the effect, especially for participants in positions with fewer or more neighbours. Interestingly, the effect persists for those with an intermediate number of neighbours in both environments.

The remainder of the paper is structured as follows: Section \ref{design} describes the experimental design and predictions, Section \ref{results} presents the results and Section \ref{conclusion} concludes.

\section{Experimental Design}\label{design}

The experiment aims at testing whether the decoy effect can be used to promote cooperation in social networks. In our benchmark network game, players are arranged in a fixed network and are exposed to the risk of suffering a shock. If a player experiences a shock, she incurs a monetary loss. The risk of a player is larger the higher the number of direct neighbors in the network.\footnote{In the experiment, this is explained to participants using neutral language (see Appendix \ref{instructions}).} Specifically, each player is endowed with 150 ECU and faces the risk of losing 100 ECU. Each player must decide whether they want to buy costly protection against this risk. Importantly, the decision to buy protection generates positive externalities for other members in the network, reducing the overall spread of risk.

We implement four experimental treatments that vary according to the presence or absence of a decoy protection option and network externalities. The \textit{Baseline Individual} treatment includes neither externalities nor a decoy. The \textit{Baseline Network} treatment introduces externalities but no  decoy. The \textit{Decoy Individual} treatment includes a decoy but no externalities. The \textit{Decoy Network} treatment combines both externalities and the presence of a decoy.

\begin{figure}[H]
\centering
\includegraphics[width=0.7\textwidth]{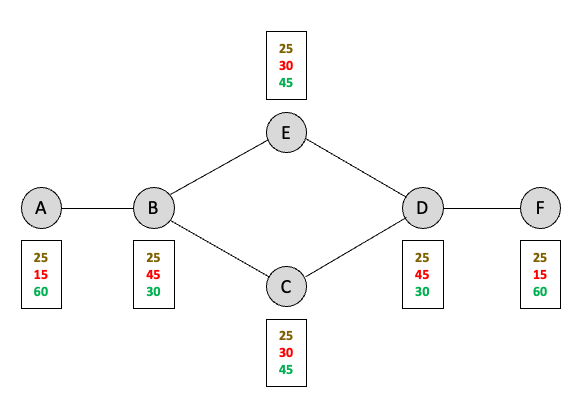}
\caption{Initial network}
\label{fig:Initial network}
\end{figure}

\paragraph{Baseline Individual treatment} Every participant has a box with 100 balls, that can be either green, brown or red. At the end of each round, one ball is drawn from each participant's box. If the ball drawn is either red or brown, the participant loses 100 ECU. If the ball drawn is green, the participant does not lose anything. The balls are drawn independently for each participant. Participants are randomly assigned to one of six positions in the network (A to F), which differ in their initial risk exposure. Positions A and F face the lowest risk, with 15 red balls, 25 brown balls, and 60 green balls, corresponding to a $40\%$ chance of losing 100 ECU. Positions C and E face a medium level of risk, with 30 red balls, 25 brown balls, and 45 green balls, corresponding to a $55\%$ chance of losing 100 ECU. Positions B and D face the highest risk, with 45 red balls, 25 brown balls, and 30 green balls, resulting in a $70\%$ chance of losing 100 ECU. Figure~\ref{fig:Initial network} illustrates the initial risk situation in the network.

Participants must decide whether or not to purchase protection against risk. In the experiment, participants purchase protection by buying Token X at a cost of 32 ECU. Token X lowers risk by replacing a given number of red balls with an equal number of green balls, depending on the participant’s network position: (i) in positions A and F, 10 red balls are replaced with 10 green balls; (ii) in positions C and E, 20 red balls are replaced with 20 green balls; and (iii) in positions B and D, 30 red balls are replaced with 30 green balls.

\paragraph{Baseline Network Treatment}
This treatment is identical to the Baseline Individual treatment, with one key difference: the decision to buy Token X affects not only the participant’s own risk but also the risk faced by others in the network. Specifically, when a participant buys Token X, (i) a given number of their own red balls are replaced with green balls, following the same rule as in the Baseline Individual treatment; (ii) 5 brown balls in the box of every other group member are replaced with green balls; and (iii) for each neighbour who did not buy any token, 10 of their red balls are also replaced with green balls. Thus, individual choices generate positive externalities: buying protection reduces not only one's own risk, but also that of others by directly altering the composition of balls in their boxes.

\paragraph{Decoy Individual Treatment}
This treatment extends the Baseline Individual treatment by introducing a second protection option: Token Y. Participants now choose between three options: no protection, buying Token X, or buying Token Y. Token X produces the same effect as in the Baseline Individual treatment: it costs 32~ECU and replaces 10, 20, or 30 red balls with green balls, depending on the participant’s position in the network. Token Y is a strictly dominated alternative, as it is more expensive and less effective: it costs 42~ECU and replaces only 8, 16, or 24 red balls with green balls, again depending on the participant's position. As in the Baseline Individual treatment, there are no externalities: each participant’s decision affects only their own risk.

\paragraph{Decoy Network Treatment}
This treatment combines the externalities of the network setup with the presence of a decoy. As in the Decoy Individual treatment, participants choose between three options: not buying any protection, buying Token X (at 32~ECU), or buying Token Y (at 42~ECU). Token X has the same effect as in the Baseline Network treatment, and additionally replaces 2 red balls with green balls for each neighbour who buys Token Y. Token Y is less effective and more expensive; it replaces 5 brown balls with green balls for all group members, plus 8 additional red balls with green balls for each non-buying neighbour. This treatment allows us to examine whether the decoy effect persists in strategic environments where individual decisions have consequences for others.

\paragraph{Implementation}
Each session consists of two treatments, with 10 rounds per treatment. After each round, participants are shown the decisions and risk levels of all group members. We run four types of sessions: (i) Individual Baseline followed by Network Baseline (2 sessions with 48 subjects in total), (ii) Network Baseline followed by Individual Baseline  (7 sessions with 102 subjects in total), (iii) Individual Decoy followed by Network Decoy  (3 sessions with 48 subjects in total), and (iv) Network Decoy followed by Individual Decoy  (6 sessions with 90 subjects in total) . One round per treatment is randomly selected for payment, and participants are paid at the end of the experiment. Payoffs depend on whether protection was purchased and on the outcome of the ball draw (green resulting in no loss, red or brown resulting in a 100 ECU loss).

\paragraph{Procedure} All our experimental sessions were conducted at the laboratory of the Nuffield Centre for Experimental Social Sciences (CESS). All participants were recruited from the participant pool at the CESS. The experiment was computerised and programmed using the experimental software o-Tree \cite{chen}.\footnote{The experiment was pre-registered on AsPredicted (\#136916).}

\section{Results}\label{results}

\subsection{Aggregate analysis}

\paragraph{Decoy effect}
We begin by analysing the effect of the decoy, specifically whether its presence reduces the frequency of choosing the competitor option (not buying protection) and increases the choice of the target option (token X) in both individual and network environments.
Table~\ref{tab:token_decisions_percent_only} presents the distribution of choices -- no protection, Token X, or Token Y -- by treatment and by the order in which the treatment appeared in the session (i.e., periods 1–10 vs. 11–20). The top panel of the table also reports decisions made in the very first round. All the relevant comparisons and tests are reported in Table~\ref{tab:test}.

Across both individual and network environments, the differences between the decoy treatments and their respective baselines consistently go in the expected direction: the presence of a decoy reduces the frequency of the no-buy decision. In the network environment, these differences are statistically significant based on one-tailed \textit{t}-tests: in all rounds combined ($p = 0.0426$) and in the first 10 rounds ($p = 0.0812$).

When examining the decoy effect in terms of increased selection of the target option (Token X), the results are puzzling. In the individual environment, all differences between decoy and baseline treatments go in the expected direction, i.e., the frequency of buying Token X is higher under the decoy treatment, but they fail to reach statistical significance. In contrast, in the network environment we observe a reduction in the frequency of choosing Token X in the presence of a decoy, with the exception of rounds 11--20. None of these differences are statistically significant, except in round 1, where the effect is significant at the 5\% level ($p = 0.0309$, one-tailed \textit{t}-test). This pattern may be driven by a higher share of participants selecting the dominated option (Token Y) in the early periods, particularly in the network environment.

\renewcommand{\arraystretch}{0.85} % Riduce l'interlinea
\begin{table}[H]
\centering
\caption{Token decisions by part and treatment (row percentages)}
\begin{tabular}{llccc}
\toprule
 & \textbf{Treatment} & \textbf{No Buy} & \textbf{Token X} & \textbf{Token Y} \\
\midrule
\multirow{4}{*}{Round 1} 
  & bas-ind & 56.3\% & 43.7\% & -- \\
  & bas-net & 43.1\% & 56.9\% & -- \\
  & dec-ind & 47.9\% & 47.9\% & 4.2\% \\
  & dec-net & 40.0\% & 43.3\% & 16.7\% \\
\midrule
\multirow{4}{*}{Rounds 1--10} 
  & bas-ind & 54.4\% & 45.6\% & -- \\
  & bas-net & 53.7\% & 46.3\% & -- \\
  & dec-ind & 46.9\% & 49.4\% & 3.8\% \\
  & dec-net & 45.4\% & 43.7\% & 10.9\% \\
\midrule
\multirow{4}{*}{Rounds 11--20} 
  & bas-ind & 37.2\% & 62.8\% & -- \\
  & bas-net & 54.2\% & 45.8\% & -- \\
  & dec-ind & 29.3\% & 63.6\% & 7.1\% \\
  & dec-net & 46.7\% & 47.7\% & 5.6\% \\
\midrule
\multirow{4}{*}{All Rounds} 
  & bas-ind & 42.7\% & 57.3\% & -- \\
  & bas-net & 53.9\% & 46.1\% & -- \\
  & dec-ind & 35.4\% & 58.6\% & 5.9\% \\
  & dec-net & 45.9\% & 45.1\% & 9.1\% \\
\bottomrule
\end{tabular}
\label{tab:token_decisions_percent_only}
\end{table}

\paragraph{Network environment versus individual environment} We now turn to the effect of network environments as compared to individual environments,  with and without the presence of a decoy. 
In network environments, we observe that in the early periods participants are more likely to purchase protection (either token) with respect to individual environments. When a decoy is available, the frequency of choosing Token X decreases, but this reduction is more than offset by the increased selection of the dominated option, Token Y. 

In contrast, during periods 11--20, participants in network environments are less likely to buy any token compared to individual treatments. Both the share of participants choosing Token X and the overall purchase rate are lower in network treatments than in the individuals treatments (the differences are about $16\%$--$17\%$). 

These differences are statistically significant at the 5\% level: for the comparison of the no-buy decision, the $p$-values are 0.0367 and 0.0181 without and with the decoy, respectively; for the comparison of Token X purchases, in treatments with decoy the corresponding $p$-value is 0.0301.

\begin{figure}[H]
    \centering
    \includegraphics[width=1\textwidth]{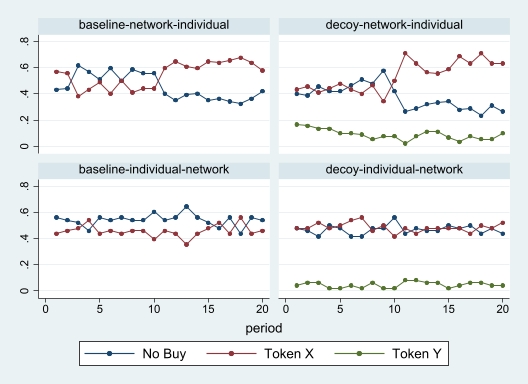}
    \caption{Average frequencies of decisions by treatment and period}
    \label{fig:graph}
\end{figure}

\paragraph{Order effects} We find that treatment ordering plays a role in shaping behaviour. Specifically, when individual treatments are conducted in the second part of a session (after participants have experienced a network treatment), participants tend to purchase protection considerably more often (see Figure \ref{fig:graph}). The frequency of selecting Token X is also higher compared to when these treatments are run in the first part. This effect is statistically significant: in the \textit{Baseline Individual} and \textit{Decoy Individual} conditions, the share of protection choices increases by approximately $17\%$ when these treatments follow a network treatment, with \textit{p}-values of 0.0413 and 0.0141, respectively. 

These findings suggest that prior exposure to strategic environments with externalities can influence subsequent behaviour, even when such externalities are no longer present. Finally, when we compare individual treatments played in the second part with network treatments played in the first part, we find an increase in both overall protection and Token X choices, ranging between $16\%$ and $20\%$, with statistical significance at the $1\%$ level.

\begin{table}[H]
\centering
\caption{Treatment comparisons}
\label{tab:test}
\resizebox{\textwidth}{!}{%
\begin{threeparttable}
\begin{tabular}{lllcccccc}
\multicolumn{9}{l}{\textbf{Baseline vs. Decoy}} \\ \hline
Period & Condition & Variable & Baseline & Decoy & diff & $p_{\text{2sided}}$ & $p_{\text{left}}$ & $p_{\text{right}}$ \\ \hline
Period 1 & Individual & No Buy   & 0.563 & 0.479 &  0.084 & 0.419 & 0.790 & 0.210 \\
Period 1 & Individual & Token X  & 0.438 & 0.479 & -0.041 & 0.686 & 0.343 & 0.657 \\
Period 1 & Network    & No Buy   & 0.431 & 0.400 &  0.031 & 0.662 & 0.669 & 0.331 \\
Period 1 & Network    & Token X  & 0.569 & 0.433 &  0.136 & 0.062 & 0.969 & 0.031 \\
Part 1   & Individual & No Buy   & 0.544 & 0.469 &  0.075 & 0.369 & 0.815 & 0.185 \\
Part 1   & Individual & Token X  & 0.456 & 0.494 & -0.038 & 0.655 & 0.327 & 0.673 \\
Part 1   & Network    & No Buy   & 0.537 & 0.454 &  0.083 & 0.162 & 0.919 & 0.081 \\
Part 1   & Network    & Token X  & 0.463 & 0.437 &  0.026 & 0.637 & 0.681 & 0.319 \\
Part 2   & Individual & No Buy   & 0.372 & 0.293 &  0.079 & 0.212 & 0.894 & 0.106 \\
Part 2   & Individual & Token X  & 0.628 & 0.636 & -0.008 & 0.907 & 0.454 & 0.546 \\
Part 2   & Network    & No Buy   & 0.542 & 0.467 &  0.075 & 0.351 & 0.825 & 0.175 \\
Part 2   & Network    & Token X  & 0.458 & 0.477 & -0.019 & 0.821 & 0.410 & 0.590 \\
All      & Individual & No Buy   & 0.427 & 0.354 &  0.073 & 0.184 & 0.908 & 0.092 \\
All      & Individual & Token X  & 0.573 & 0.586 & -0.013 & 0.807 & 0.404 & 0.596 \\
All      & Network    & No Buy   & 0.539 & 0.459 &  0.080 & 0.085 & 0.957 & 0.043 \\
All      & Network    & Token X  & 0.461 & 0.451 &  0.010 & 0.813 & 0.594 & 0.406 \\
\hline \\

\multicolumn{9}{l}{\textbf{Individual vs. Network}} \\ \hline
Period & Condition & Variable & Individual & Network & diff & $p_{\text{all}}$ & $p_l$ & $p_u$ \\ \hline
Period 1 & Baseline & No Buy   & 0.563 & 0.431 &  0.132 & 0.135 & 0.932 & 0.068 \\
Period 1 & Decoy    & No Buy   & 0.479 & 0.400 &  0.079 & 0.374 & 0.813 & 0.187 \\
Period 1 & Decoy    & Token X  & 0.479 & 0.433 &  0.046 & 0.609 & 0.695 & 0.305 \\
Part 1   & Baseline & No Buy   & 0.544 & 0.537 &  0.007 & 0.918 & 0.541 & 0.459 \\
Part 1   & Decoy    & No Buy   & 0.469 & 0.454 &  0.015 & 0.857 & 0.572 & 0.428 \\
Part 1   & Decoy    & Token X  & 0.494 & 0.437 &  0.057 & 0.450 & 0.775 & 0.225 \\
Part 2   & Baseline & No Buy   & 0.372 & 0.542 & -0.170 & 0.037 & 0.018 & 0.982 \\
Part 2   & Decoy    & No Buy   & 0.293 & 0.467 & -0.174 & 0.018 & 0.009 & 0.991 \\
Part 2   & Decoy    & Token X  & 0.636 & 0.477 &  0.159 & 0.030 & 0.985 & 0.015 \\
\hline \\

\multicolumn{9}{l}{\textbf{Part 1 vs. Part 2 in the same session}} \\ \hline
Condition& Session type & Variable & Part 1& Part 2 & diff & $p_{\text{all}}$ & $p_l$ & $p_u$ \\ \hline
Baseline& Individual$\to$Network & No Buy   & 0.544 & 0.542 &  0.002 & 0.966 & 0.517 & 0.483 \\
Baseline& Network$\to$Individual & No Buy   & 0.537 & 0.372 &  0.165 & 0.000 & 1.000 & 0.000 \\
Decoy& Individual$\to$Network & No Buy   & 0.469 & 0.467 &  0.002 & 0.983 & 0.508 & 0.492 \\
Decoy& Individual$\to$Network & Token X  & 0.494 & 0.477 &  0.017 & 0.883 & 0.559 & 0.441 \\
Decoy& Network$\to$Individual & No Buy   & 0.454 & 0.293 &  0.161 & 0.003 & 0.999 & 0.001 \\
Decoy& Network$\to$Individual & Token X  & 0.437 & 0.636 & -0.199 & 0.002 & 0.001 & 0.999 \\
\hline \\

\multicolumn{9}{l}{\textbf{Part 1 vs. Part 2 in the same treatment}} \\ \hline
 & Treatment & Variable & Part 1 & Part 2 & diff & $p_{\text{all}}$ & $p_l$ & $p_u$ \\ \hline
& Baseline Individual & No Buy   & 0.544 & 0.372 &  0.172 & 0.041 & 0.979 & 0.021 \\
& Baseline Network    & No Buy   & 0.537 & 0.542 & -0.005 & 0.941 & 0.470 & 0.530 \\
& Decoy Individual    & No Buy   & 0.469 & 0.293 &  0.176 & 0.014 & 0.993 & 0.007 \\
& Decoy Individual    & Token X  & 0.494 & 0.636 & -0.142 & 0.040 & 0.020 & 0.980 \\
& Decoy Network       & No Buy   & 0.454 & 0.467 & -0.013 & 0.880 & 0.440 & 0.560 \\
& Decoy Network       & Token X  & 0.437 & 0.477 & -0.040 & 0.606 & 0.303 & 0.697 \\
\hline
\end{tabular}
\end{threeparttable}
}
\end{table}

\subsection{Analysis by degree}
Tables \ref{tab:degree1}, \ref{tab:degree2}, and \ref{tab:degree3} report the observed decisions for degrees 1, 2, and 3, respectively. 
A preliminary examination indicates that the purchase of protection -- and, more specifically, of Token X -- tends to increase with the degree across all experimental conditions.

\textbf{One link}
The behaviour of participants with one link is summarised in Table~\ref{tab:degree1}. All relevant comparisons and tests are reported in Table~\ref{tab:testdegree1}. Overall, the presence of a decoy reduces the frequency of no-buy decisions and increases the choice of Token X. These effects seem to be driven primarily by treatments conducted in the second half of the sessions (i.e., rounds 11--20). 
Most of these differences are not statistically significant, with the only exception being the reduction in the no-buy option under the decoy treatment in the individual environment, which shows a 15\% decrease ($p$-value = 0.0676).

\begin{table}[H]
\centering
\caption{Token decisions by treatment and part for degree 1 (row percentages)}
\renewcommand{\arraystretch}{0.9}
\begin{tabular}{llccc}
\toprule
 & \textbf{Treatment} & \textbf{No Buy} & \textbf{Token X} & \textbf{Token Y} \\
\midrule
\multirow{4}{*}{Round 1} 
& bas-ind  & 81.3\% & 18.8\% & -- \\
& bas-net  & 64.7\% & 35.3\% & -- \\
& dec-ind  & 75.0\% & 25.0\% & -- \\
& dec-net  & 50.0\% & 43.3\% & 6.7\% \\
\midrule
\multirow{4}{*}{Rounds 1--10} 
  & bas-ind  & 78.8\% & 21.3\% & -- \\
  & bas-net  & 65.0\% & 35.0\% & -- \\
  & dec-ind  & 73.8\% & 25.6\% & 0.6\% \\
  & dec-net  & 56.0\% & 37.3\% & 6.7\% \\
\midrule
\multirow{4}{*}{Rounds 11--20} 
  & bas-ind  & 51.8\% & 48.2\% & -- \\
  & bas-net  & 75.0\% & 25.0\% & -- \\
  & dec-ind  & 36.0\% & 58.3\% & 5.7\% \\
  & dec-net  & 63.8\% & 36.2\% & -- \\
\midrule
\multirow{4}{*}{All Rounds} 
  & bas-ind  & 60.4\% & 39.6\% & -- \\
  & bas-net  & 68.2\% & 31.8\% & -- \\
  & dec-ind  & 49.1\% & 47.0\% & 3.9\% \\
  & dec-net  & 58.7\% & 37.0\% & 4.3\% \\
\bottomrule
\end{tabular}
\label{tab:degree1}
\end{table}

The impact of the network environment compared to the individual one depends on the timing of the treatment within the session. When comparing treatments played first (rounds 1--10), the network environment reduces the frequency of no-buy decisions and increases the selection of Token X relative to the individual environment. However, when treatments are played second (rounds 11--20), the opposite occurs: the individual environments generate higher protection rates than the network ones (in the baseline $+23\%$ with $p$-value=0.0811; when the decoy is present $+28\%$ with $p$-value=0.0581).

\begin{table}[H]
\centering
\caption{Treatment comparisons for degree 1}
\label{tab:testdegree1}
\resizebox{\textwidth}{!}{%
\begin{threeparttable}
\begin{tabular}{lllcccccc}
\multicolumn{9}{l}{\textbf{Baseline vs. Decoy}} \\ \hline
Period & Condition & Variable & Baseline & Decoy & diff & $p_{\text{2sided}}$ & $p_{\text{left}}$ & $p_{\text{right}}$ \\ \hline
Period 1 & Individual & No Buy   & 0.813 & 0.750 &  0.063 & 0.681 & 0.660 & 0.340 \\
Period 1 & Individual & Token X  & 0.188 & 0.250 & -0.062 & 0.681 & 0.340 & 0.660 \\
Period 1 & Network    & No Buy   & 0.647 & 0.500 &  0.147 & 0.241 & 0.879 & 0.121 \\
Period 1 & Network    & Token X  & 0.353 & 0.433 & -0.080 & 0.518 & 0.259 & 0.741 \\
Part 1   & Individual & No Buy   & 0.788 & 0.737 &  0.051 & 0.675 & 0.663 & 0.337 \\
Part 1   & Individual & Token X  & 0.213 & 0.256 & -0.043 & 0.703 & 0.351 & 0.649 \\
Part 1   & Network    & No Buy   & 0.650 & 0.560 &  0.090 & 0.399 & 0.800 & 0.200 \\
Part 1   & Network    & Token X  & 0.350 & 0.373 & -0.023 & 0.820 & 0.410 & 0.590 \\
Part 2   & Individual & No Buy   & 0.518 & 0.360 &  0.158 & 0.135 & 0.932 & 0.068 \\
Part 2   & Individual & Token X  & 0.482 & 0.583 & -0.101 & 0.347 & 0.173 & 0.827 \\
Part 2   & Network    & No Buy   & 0.750 & 0.638 &  0.112 & 0.518 & 0.741 & 0.259 \\
Part 2   & Network    & Token X  & 0.250 & 0.362 & -0.112 & 0.518 & 0.259 & 0.741 \\
All      & Individual & No Buy   & 0.604 & 0.491 &  0.113 & 0.215 & 0.893 & 0.107 \\
All      & Individual & Token X  & 0.396 & 0.470 & -0.074 & 0.411 & 0.205 & 0.795 \\
All      & Network    & No Buy   & 0.682 & 0.587 &  0.095 & 0.290 & 0.855 & 0.145 \\
All      & Network    & Token X  & 0.318 & 0.370 & -0.052 & 0.554 & 0.277 & 0.723 \\
\hline \\

\multicolumn{9}{l}{\textbf{Individual vs. Network}} \\ \hline
Period & Condition & Variable & Individual & Network & diff & $p_{\text{all}}$ & $p_l$ & $p_u$ \\ \hline
Period 1 & Baseline & No Buy   & 0.813 & 0.647 &  0.166 & 0.242 & 0.879 & 0.121 \\
Period 1 & Decoy    & No Buy   & 0.750 & 0.500 &  0.250 & 0.105 & 0.947 & 0.053 \\
Period 1 & Decoy    & Token X  & 0.250 & 0.433 & -0.183 & 0.229 & 0.114 & 0.886 \\
Part 1   & Baseline & No Buy   & 0.788 & 0.650 &  0.138 & 0.244 & 0.878 & 0.122 \\
Part 1   & Decoy    & No Buy   & 0.737 & 0.560 &  0.177 & 0.175 & 0.912 & 0.088 \\
Part 1   & Decoy    & Token X  & 0.256 & 0.373 & -0.117 & 0.331 & 0.165 & 0.835 \\
Part 2   & Baseline & No Buy   & 0.518 & 0.750 & -0.232 & 0.081 & 0.041 & 0.959 \\
Part 2   & Decoy    & No Buy   & 0.360 & 0.638 & -0.278 & 0.058 & 0.029 & 0.971 \\
Part 2   & Decoy    & Token X  & 0.583 & 0.362 &  0.221 & 0.138 & 0.931 & 0.069 \\
\hline \\

\multicolumn{9}{l}{\textbf{Part 1 vs. Part 2 in the same session}} \\ \hline
Condition & Session type & Variable & Part 1 & Part 2 & diff & $p_{\text{all}}$ & $p_l$ & $p_u$ \\ \hline
 Baseline& Individual$\to$Network & No Buy   & 0.788 & 0.750 &  0.038 & 0.621 & 0.690 & 0.310 \\
 Baseline& Network$\to$Individual & No Buy   & 0.650 & 0.518 &  0.132 & 0.008 & 0.996 & 0.004 \\
 Decoy& Individual$\to$Network & No Buy   & 0.737 & 0.638 &  0.099 & 0.404 & 0.798 & 0.202 \\
 Decoy& Individual$\to$Network & Token X  & 0.256 & 0.362 & -0.106 & 0.376 & 0.188 & 0.812 \\
Decoy & Network$\to$Individual & No Buy   & 0.560 & 0.360 &  0.200 & 0.012 & 0.994 & 0.006 \\
Decoy & Network$\to$Individual & Token X  & 0.373 & 0.583 & -0.210 & 0.009 & 0.004 & 0.996 \\
\hline \\

\multicolumn{9}{l}{\textbf{Part 1 vs. Part 2 in the same treatment}} \\ \hline
 & Treatment & Variable & Part 1 & Part 2 & diff & $p_{\text{all}}$ & $p_l$ & $p_u$ \\ \hline
 & Baseline Individual & No Buy   & 0.788 & 0.518 &  0.270 & 0.022 & 0.989 & 0.011 \\
 & Baseline Network    & No Buy   & 0.650 & 0.750 & -0.100 & 0.455 & 0.228 & 0.772 \\
 & Decoy Individual    & No Buy   & 0.737 & 0.360 &  0.377 & 0.007 & 0.996 & 0.004 \\
 & Decoy Individual    & Token X  & 0.256 & 0.583 & -0.327 & 0.020 & 0.010 & 0.990 \\
& Decoy Network       & No Buy   & 0.560 & 0.638 & -0.078 & 0.580 & 0.290 & 0.710 \\
 & Decoy Network       & Token X  & 0.373 & 0.362 &  0.011 & 0.935 & 0.532 & 0.468 \\
\hline
\end{tabular}
\end{threeparttable}
}
\end{table}

Regarding the ordering effect, we observe a similar pattern as in the aggregate analysis: participants are more likely to buy protection in individual treatments when these are run in the second part of the session, compared to when the same treatment is run in the first part. The no buy option is-27$\%$ with $p$-value=0.0221 in \emph{Baseline Individual}, and -38$\%$ $p$-value=0.0073 in \emph{Decoy Individual}). The purchase of Token X is +33$\%$ $p$-value=0.0203 in \emph{Decoy Individual}.

In contrast, in the network treatments, we observe the opposite trend: participants purchase less protection when the treatment is run in the second part, compared to the same treatment run in the first part. However, these differences are not significant.

\paragraph{Two links}
The behaviour of participants with two links is summarised in Table~\ref{tab:degree2}. All the relevant comparisons and tests are reported in Table~\ref{tab:testdegree2}. 
Overall, the presence of a decoy reduces the frequency of no-buy decisions especially in treatments with network externalities (-11$\%$ $p$-value=0.0630). This effect seems to be driven primarily by treatments in the network condition conducted in the second half of the session (i.e., rounds 11--20).  Here we observe a 22$\%$ reduction of the no buy option ($p$-value=0.0363).

The impact of the network environment compared to the individual one results in an increase of the frequency of no-buy decisions and a decrease in the frequency of Token X choices. Almost all differences are not significant with the exception of the baseline treatments in periods 11--20 (no-buy option is +23$\%$ with $p$-value=0.0863).

Regarding the order effect, we do not observe significant differences between running treatments in the first or in the second part of the session. The overall pattern is similar as in the aggregate, except for \emph{Decoy Network}: when it is run in the second part, the overall purchase of protection and the purchase of Token X increase.

\begin{table}[H]
\centering
\caption{Token decisions by treatment and part for degree 2 (row percentages)}
\renewcommand{\arraystretch}{0.9}
\begin{tabular}{llccc}
\toprule
 & \textbf{Treatment} & \textbf{No Buy} & \textbf{Token X} & \textbf{Token Y} \\
\midrule
\multirow{4}{*}{Round 1} 
& bas-ind  & 50.0\% & 50.0\% & -- \\
& bas-net  & 50.0\% & 50.0\% & -- \\
& dec-ind  & 50.0\% & 43.8\% & 6.2\% \\
& dec-net  & 33.3\% & 50.0\% & 16.7\% \\
\midrule
\multirow{4}{*}{Rounds 1 -- 10} 
& bas-ind  & 48.1\% & 51.9\% & -- \\
& bas-net  & 52.4\% & 47.6\% & -- \\
& dec-ind  & 40.0\% & 53.1\% & 6.9\% \\
& dec-net  & 46.0\% & 43.3\% & 10.7\% \\
\midrule
\multirow{4}{*}{Rounds 11-- 20}
& bas-ind  & 38.2\% & 61.8\% & -- \\
& bas-net  & 61.3\% & 38.8\% & -- \\
& dec-ind  & 34.0\% & 58.7\% & 7.3\% \\
& dec-net  & 39.4\% & 51.9\% & 8.8\% \\
\midrule
\multirow{4}{*}{All Rounds}
& bas-ind  & 41.4\% & 58.6\% & -- \\
& bas-net  & 55.2\% & 44.8\% & -- \\
& dec-ind  & 36.1\% & 56.7\% & 7.2\% \\
& dec-net  & 43.7\% & 46.3\% & 10.0\% \\
\bottomrule
\end{tabular}
\label{tab:degree2}
\end{table}

\begin{table}[H]
\centering
\caption{Treatment comparisons for degree 2}
\label{tab:testdegree2}
\resizebox{\textwidth}{!}{%
\begin{threeparttable}
\begin{tabular}{lllcccccc}
\multicolumn{9}{l}{\textbf{Baseline vs. Decoy}} \\ \hline
Period & Condition & Variable & Baseline & Decoy & diff & $p_{\text{2sided}}$ & $p_{\text{left}}$ & $p_{\text{right}}$ \\ \hline
Period 1 & Individual & No Buy   & 0.500 & 0.500 &  0.000 & 1.000 & 0.500 & 0.500 \\
Period 1 & Individual & Token X  & 0.500 & 0.438 &  0.062 & 0.733 & 0.633 & 0.367 \\
Period 1 & Network    & No Buy   & 0.500 & 0.333 &  0.167 & 0.183 & 0.908 & 0.092 \\
Period 1 & Network    & Token X  & 0.500 & 0.500 &  0.000 & 1.000 & 0.500 & 0.500 \\
Part 1   & Individual & No Buy   & 0.481 & 0.400 &  0.081 & 0.458 & 0.771 & 0.229 \\
Part 1   & Individual & Token X  & 0.519 & 0.531 & -0.012 & 0.910 & 0.455 & 0.545 \\
Part 1   & Network    & No Buy   & 0.524 & 0.460 &  0.064 & 0.515 & 0.742 & 0.258 \\
Part 1   & Network    & Token X  & 0.476 & 0.433 &  0.043 & 0.614 & 0.693 & 0.307 \\
Part 2   & Individual & No Buy   & 0.382 & 0.340 &  0.042 & 0.690 & 0.655 & 0.345 \\
Part 2   & Individual & Token X  & 0.618 & 0.587 &  0.031 & 0.764 & 0.618 & 0.382 \\
Part 2   & Network    & No Buy   & 0.612 & 0.394 &  0.218 & 0.073 & 0.964 & 0.036 \\
Part 2   & Network    & Token X  & 0.387 & 0.519 & -0.132 & 0.354 & 0.177 & 0.823 \\
All      & Individual & No Buy   & 0.414 & 0.361 &  0.053 & 0.499 & 0.751 & 0.249 \\
All      & Individual & Token X  & 0.586 & 0.567 &  0.019 & 0.809 & 0.596 & 0.404 \\
All      & Network    & No Buy   & 0.552 & 0.437 &  0.115 & 0.126 & 0.937 & 0.063 \\
All      & Network    & Token X  & 0.448 & 0.463 & -0.015 & 0.835 & 0.418 & 0.582 \\
\hline \\

\multicolumn{9}{l}{\textbf{Individual vs. Network}} \\ \hline
Period & Condition & Variable & Individual & Network & diff & $p_{\text{all}}$ & $p_l$ & $p_u$ \\ \hline
Period 1 & Baseline & No Buy   & 0.500 & 0.500 &  0.000 & 1.000 & 0.500 & 0.500 \\
Period 1 & Decoy    & No Buy   & 0.500 & 0.333 &  0.167 & 0.280 & 0.860 & 0.140 \\
Period 1 & Decoy    & Token X  & 0.438 & 0.500 & -0.062 & 0.694 & 0.347 & 0.653 \\
Part 1   & Baseline & No Buy   & 0.481 & 0.524 & -0.043 & 0.690 & 0.345 & 0.655 \\
Part 1   & Decoy    & No Buy   & 0.400 & 0.460 & -0.060 & 0.612 & 0.306 & 0.694 \\
Part 1   & Decoy    & Token X  & 0.531 & 0.433 &  0.098 & 0.318 & 0.841 & 0.159 \\
Part 2   & Baseline & No Buy   & 0.382 & 0.612 & -0.230 & 0.086 & 0.043 & 0.957 \\
Part 2   & Decoy    & No Buy   & 0.340 & 0.394 & -0.054 & 0.625 & 0.312 & 0.688 \\
Part 2   & Decoy    & Token X  & 0.587 & 0.519 &  0.068 & 0.565 & 0.718 & 0.282 \\
\hline \\

\multicolumn{9}{l}{\textbf{Part 1 vs. Part 2 in the same session}} \\ \hline
 Condition & Session type & Variable & Part 1 & Part 2 & diff & $p_{\text{all}}$ & $p_l$ & $p_u$ \\ \hline
 Baseline& Individual$\to$Network & No Buy   & 0.481 & 0.612 & -0.131 & 0.164 & 0.082 & 0.918 \\
 Baseline& Network$\to$Individual & No Buy   & 0.524 & 0.382 &  0.142 & 0.005 & 0.997 & 0.003 \\
 Decoy& Individual$\to$Network & No Buy   & 0.400 & 0.394 &  0.006 & 0.954 & 0.523 & 0.477 \\
 Decoy& Individual$\to$Network & Token X  & 0.531 & 0.519 &  0.012 & 0.930 & 0.535 & 0.465 \\
 Decoy& Network$\to$Individual & No Buy   & 0.460 & 0.340 &  0.120 & 0.164 & 0.918 & 0.082 \\
 Decoy& Network$\to$Individual & Token X  & 0.433 & 0.587 & -0.154 & 0.036 & 0.018 & 0.982 \\
\hline \\

\multicolumn{9}{l}{\textbf{Part 1 vs. Part 2 in the same treatment}} \\ \hline
 & Treatment& Variable & Part 1 & Part 2 & diff & $p_{\text{all}}$ & $p_l$ & $p_u$ \\ \hline
 & Baseline Individual & No Buy   & 0.481 & 0.382 &  0.099 & 0.458 & 0.771 & 0.229 \\
 & Baseline Network    & No Buy   & 0.524 & 0.612 & -0.088 & 0.390 & 0.195 & 0.805 \\
 & Decoy Individual    & No Buy   & 0.400 & 0.340 &  0.060 & 0.560 & 0.720 & 0.280 \\
 & Decoy Individual    & Token X  & 0.531 & 0.587 & -0.056 & 0.569 & 0.285 & 0.715 \\
 & Decoy Network       & No Buy   & 0.460 & 0.394 &  0.066 & 0.595 & 0.703 & 0.297 \\
 & Decoy Network       & Token X  & 0.433 & 0.519 & -0.086 & 0.469 & 0.235 & 0.765 \\
\hline
\end{tabular}
\end{threeparttable}
}
\end{table}

\paragraph{Three links} The behaviour of the participants with three links is summarised in Table~\ref{tab:degree3}. All the relevant comparisons and tests are reported in Table~\ref{tab:testdegree3}. 

Overall, the decoy appears to have little effect. 
An exception is the observed reduction in the purchase of Token~X when the decoy is introduced in the network environment ($-10\%$, $p = 0.094$). 
However, when disaggregating by period, we find that in the individual environment the introduction of the decoy leads to changes in the purchase of protection in the expected direction, although the effects are not statistically significant. In the network environment the changes are in the opposite direction and are significant in period~1 (No Buy: $+22\%$, $p = 0.022$; Token~X: $-49\%$, $p < 0.001$).

Comparing network treatments with individual treatments, we observe that in the presence of the decoy, the introduction of network externalities increases the frequency of the no-buy option and reduces the purchase of Token~X. 
The latter effect is significant in period~1 and in part~1, with reductions of $38\%$ ($p = 0.013$) and $19\%$ ($p = 0.033$), respectively. 

We do not observe relevant or significant order effects, i.e., differences arising from running treatments in the first versus the second part of the session.

\begin{table}[H]
\centering
\caption{Token decisions by treatment and part for degree 3 (row percentages)}
\renewcommand{\arraystretch}{0.9}
\begin{tabular}{llccc}
\toprule
 & \textbf{Treatment} & \textbf{No Buy} & \textbf{Token X} & \textbf{Token Y} \\
\midrule
\multirow{4}{*}{Round 1} 
& bas-ind  & 37.5\% & 62.5\% & -- \\
& bas-net  & 14.7\% & 85.3\% & -- \\
& dec-ind  & 18.8\% & 75.0\% & 6.2\% \\
& dec-net  & 36.7\% & 36.7\% & 26.7\% \\
\midrule
\multirow{4}{*}{Rounds  1 -- 10} 
& bas-ind  & 36.3\% & 63.8\% & -- \\
& bas-net  & 43.8\% & 56.2\% & -- \\
& dec-ind  & 26.9\% & 69.4\% & 3.8\% \\
& dec-net  & 34.3\% & 50.3\% & 15.3\% \\
\midrule
\multirow{4}{*}{Round 11 -- 20} 
& bas-ind  & 21.5\% & 78.5\% & -- \\
& bas-net  & 26.3\% & 73.8\% & -- \\
& dec-ind  & 18.0\% & 73.7\% & 8.3\% \\
& dec-net  & 36.9\% & 55.0\% & 8.1\% \\
\midrule
\multirow{4}{*}{All Rounds} 
& bas-ind  & 26.2\% & 73.8\% & -- \\
& bas-net  & 38.2\% & 61.8\% & -- \\
& dec-ind  & 21.1\% & 72.2\% & 6.7\% \\
& dec-net  & 35.2\% & 52.0\% & 12.8\% \\
\bottomrule
\end{tabular}
\label{tab:degree3}
\end{table}

\begin{table}[H]
\centering
\caption{Treatment comparisons for degree 3}
\label{tab:testdegree3}
\resizebox{\textwidth}{!}{%
\begin{threeparttable}
\begin{tabular}{lllcccccc}
\multicolumn{9}{l}{\textbf{Baseline vs. Decoy}} \\ \hline
Period & Condition & Variable & Baseline & Decoy & diff & $p_{\text{2sided}}$ & $p_{\text{left}}$ & $p_{\text{right}}$ \\ \hline
Period 1 & Individual & No Buy   & 0.375 & 0.188 &  0.187 & 0.252 & 0.874 & 0.126 \\
Period 1 & Individual & Token X  & 0.625 & 0.750 & -0.125 & 0.462 & 0.231 & 0.769 \\
Period 1 & Network    & No Buy   & 0.147 & 0.367 & -0.220 & 0.044 & 0.022 & 0.978 \\
Period 1 & Network    & Token X  & 0.853 & 0.367 &  0.486 & 0.000 & 1.000 & 0.000 \\
Part 1   & Individual & No Buy   & 0.363 & 0.269 &  0.094 & 0.441 & 0.779 & 0.221 \\
Part 1   & Individual & Token X  & 0.638 & 0.694 & -0.056 & 0.645 & 0.323 & 0.677 \\
Part 1   & Network    & No Buy   & 0.438 & 0.343 &  0.095 & 0.279 & 0.860 & 0.140 \\
Part 1   & Network    & Token X  & 0.562 & 0.503 &  0.059 & 0.491 & 0.754 & 0.246 \\
Part 2   & Individual & No Buy   & 0.215 & 0.180 &  0.035 & 0.620 & 0.690 & 0.310 \\
Part 2   & Individual & Token X  & 0.785 & 0.737 &  0.048 & 0.555 & 0.723 & 0.277 \\
Part 2   & Network    & No Buy   & 0.263 & 0.369 & -0.106 & 0.469 & 0.235 & 0.765 \\
Part 2   & Network    & Token X  & 0.738 & 0.550 &  0.188 & 0.207 & 0.897 & 0.103 \\
All      & Individual & No Buy   & 0.262 & 0.211 &  0.051 & 0.411 & 0.795 & 0.205 \\
All      & Individual & Token X  & 0.738 & 0.722 &  0.016 & 0.811 & 0.595 & 0.405 \\
All      & Network    & No Buy   & 0.382 & 0.352 &  0.030 & 0.692 & 0.654 & 0.346 \\
All      & Network    & Token X  & 0.618 & 0.520 &  0.098 & 0.188 & 0.906 & 0.094 \\
\hline \\

\multicolumn{9}{l}{\textbf{Individual vs. Network}} \\ \hline
Period & Condition & Variable & Individual & Network & diff & $p_{\text{all}}$ & $p_l$ & $p_u$ \\ \hline
Period 1 & Baseline & No Buy   & 0.375 & 0.147 &  0.228 & 0.072 & 0.964 & 0.036 \\
Period 1 & Decoy    & No Buy   & 0.188 & 0.367 & -0.179 & 0.217 & 0.109 & 0.891 \\
Period 1 & Decoy    & Token X  & 0.750 & 0.367 &  0.383 & 0.013 & 0.994 & 0.006 \\
Part 1   & Baseline & No Buy   & 0.363 & 0.438 & -0.075 & 0.524 & 0.262 & 0.738 \\
Part 1   & Decoy    & No Buy   & 0.269 & 0.343 & -0.074 & 0.404 & 0.202 & 0.798 \\
Part 1   & Decoy    & Token X  & 0.694 & 0.503 &  0.191 & 0.033 & 0.983 & 0.017 \\
Part 2   & Baseline & No Buy   & 0.215 & 0.263 & -0.048 & 0.576 & 0.288 & 0.712 \\
Part 2   & Decoy    & No Buy   & 0.180 & 0.369 & -0.189 & 0.110 & 0.055 & 0.945 \\
Part 2   & Decoy    & Token X  & 0.737 & 0.550 &  0.187 & 0.163 & 0.919 & 0.081 \\
\hline \\

\multicolumn{9}{l}{\textbf{Part 1 vs. Part 2 in the same session}} \\ \hline
 Condition& Session type & Variable & Part 1 & Part 2 & diff & $p_{\text{all}}$ & $p_l$ & $p_u$ \\ \hline
Baseline & Individual$\to$Network & No Buy   & 0.363 & 0.263 &  0.100 & 0.399 & 0.800 & 0.200 \\
Baseline & Network$\to$Individual & No Buy   & 0.438 & 0.215 &  0.223 & 0.014 & 0.993 & 0.007 \\
 Decoy& Individual$\to$Network & No Buy   & 0.269 & 0.369 & -0.100 & 0.497 & 0.248 & 0.752 \\
 Decoy& Individual$\to$Network & Token X  & 0.694 & 0.550 &  0.144 & 0.380 & 0.810 & 0.190 \\
 Decoy& Network$\to$Individual & No Buy   & 0.343 & 0.180 &  0.163 & 0.042 & 0.979 & 0.021 \\
 Decoy& Network$\to$Individual & Token X  & 0.503 & 0.737 & -0.234 & 0.018 & 0.009 & 0.991 \\
\hline \\

\multicolumn{9}{l}{\textbf{Part 1 vs. Part 2 in the same treatment}} \\ \hline
 & Treatment & Variable & Part 1 & Part 2 & diff & $p_{\text{all}}$ & $p_l$ & $p_u$ \\ \hline
 & Baseline Individual & No Buy   & 0.363 & 0.215 &  0.148 & 0.157 & 0.921 & 0.079 \\
 & Baseline Network    & No Buy   & 0.438 & 0.263 &  0.175 & 0.101 & 0.949 & 0.051 \\
 & Decoy Individual    & No Buy   & 0.269 & 0.180 &  0.089 & 0.263 & 0.868 & 0.132 \\
 & Decoy Individual    & Token X  & 0.694 & 0.737 & -0.043 & 0.674 & 0.337 & 0.663 \\
 & Decoy Network       & No Buy   & 0.343 & 0.369 & -0.026 & 0.835 & 0.417 & 0.583 \\
 & Decoy Network       & Token X  & 0.503 & 0.550 & -0.047 & 0.692 & 0.346 & 0.654 \\
\hline
\end{tabular}
\end{threeparttable}
}
\end{table}

\section{Conclusion}\label{conclusion}

Our results show relatively small differences between baseline and decoy conditions, with few statistically significant effects in both the individual and network settings. This suggests that the introduction of a decoy, Token Y, consisting of a dominated protection alternative, did not consistently shift behaviour towards purchasing the dominating Token X more frequently across periods. However, we do find that the decoy increases the share of overall protection, meaning fewer participants remain without any protection. This corresponds to a decrease in the likelihood of choosing the competitor, driven by choices of both the target and the dominated option of the decoy. 

When comparing individual and network treatments, some systematic effects emerge. In particular, the network condition tends to increase the share of the no-buy option indicating some level of free riding, especially for subjects with three links.

The choice of the decoy is particularly puzzling. While its relatively low occurrence in the individual setting could be interpreted as an error,  its occurrence is consistently higher in the network setting, especially among subjects with three links in the network condition.  This points to the role of free-riding possibilities in networks: it is possible that the necessity of purchasing protection due to the high risk exposure, which allows all neighbors to free ride, is addressed by choosing a suboptimal protection option. However, we do observe a decline in decoy choices  in later periods.

In summary, these findings suggest that while network structure can meaningfully shape choice patterns, the decoy effect alone is insufficient to generate large and consistent shifts in preferences. This implies that strategic interaction within the network, rather than the presence of a dominated alternative, is the primary driver of behavioural change in this setting.

\bibliographystyle{apalike}
\bibliography{references}

\pagebreak
\appendix

\section*{Appendix}\label{APP}

\renewcommand{\thesubsection}{\Alph{subsection}}

\section{Additional tables}\label{more-tables}

\renewcommand{\arraystretch}{0.85} % Riduce l'interlinea
\begin{table}[H]
\centering
\caption{Token decisions by part and treatment}
\begin{tabular}{llrrrrr}
\toprule
 & \textbf{Treatment} & \textbf{No Buy} & \textbf{Token X} & \textbf{Token Y}& \textbf{Total}  \\
\midrule
\multirow{5}{*}{Round 1} 
  & bas-ind & 27 (56.3\%) & 21 (43.7\%) & -- & 48 \\
  & bas-net & 44 (43.1\%) & 58 (56.9\%) & -- & 102 \\
  & dec-ind & 23 (47.9\%) & 23 (47.9\%) & 2 (4.2\%) & 48 \\
  & dec-net & 36 (40.0\%) & 39 (43.3\%) & 15 (16.7\%) & 90 \\
 \midrule
\multirow{5}{*}{Rounds 1 -- 10} 
  & bas-ind & 261 (54.4\%) & 219 (45.6\%) & -- & 480 \\
  & bas-net & 548 (53.7\%) & 472 (46.3\%) & -- & 1,020 \\
  & dec-ind & 225 (46.9\%) & 237 (49.4\%) & 18 (3.8\%) & 480 \\
  & dec-net & 409 (45.4\%) & 393 (43.7\%) & 98 (10.9\%) & 900 \\
\midrule
\multirow{5}{*}{Rounds 11 -- 20} 
  & bas-ind & 379 (37.2\%) & 641 (62.8\%) & -- & 1,020 \\
  & bas-net & 260 (54.2\%) & 220 (45.8\%) & -- & 480 \\
  & dec-ind & 264 (29.3\%) & 572 (63.6\%) & 64 (7.1\%) & 900 \\
  & dec-net & 224 (46.7\%) & 229 (47.7\%) & 27 (5.6\%) & 480 \\
\midrule
\multirow{5}{*}{All Rounds} 
  & bas-ind & 640 (42.7\%) & 860 (57.3\%) & -- & 1,500 \\
  & bas-net & 808 (53.9\%) & 692 (46.1\%) & -- & 1,500 \\
  & dec-ind & 489 (35.4\%) & 809 (58.6\%) & 82 (5.9\%) & 1,380 \\
  & dec-net & 633 (45.9\%) & 622 (45.1\%) & 125 (9.1\%) & 1,380 \\
  \bottomrule
\end{tabular}
\label{tab:token_decisions_all}
\end{table}

\begin{table}[H]
\centering
\caption{Token decisions by treatment, part, and number of links (absolute frequencies and row percentages)}
\renewcommand{\arraystretch}{0.8}
\begin{tabular}{lllrrrr}
\toprule
\textbf{Part} & \textbf{Treatment} & \textbf{Links} & \textbf{No Buy} & \textbf{Token X} & \textbf{Token Y} & \textbf{Total} \\
\midrule
\multirow{12}{*}{1--10} 
 & \multirow{3}{*}{bas-ind} 
   & 1 & 126 (78.8\%) & 34 (21.3\%) & -- & 160 \\
 & & 2 & 77 (48.1\%) & 83 (51.9\%) & -- & 160 \\
 & & 3 & 58 (36.3\%) & 102 (63.8\%) & -- & 160 \\

 & \multirow{3}{*}{bas-net} 
   & 1 & 221 (65.0\%) & 119 (35.0\%) & -- & 340 \\
 & & 2 & 178 (52.4\%) & 162 (47.6\%) & -- & 340 \\
 & & 3 & 149 (43.8\%) & 191 (56.2\%) & -- & 340 \\

 & \multirow{3}{*}{dec-ind} 
   & 1 & 118 (73.8\%) & 41 (25.6\%) & 1 (0.6\%) & 160 \\
 & & 2 & 64 (40.0\%) & 85 (53.1\%) & 11 (6.9\%) & 160 \\
 & & 3 & 43 (26.9\%) & 111 (69.4\%) & 6 (3.8\%) & 160 \\

 & \multirow{3}{*}{dec-net} 
   & 1 & 168 (56.0\%) & 112 (37.3\%) & 20 (6.7\%) & 300 \\
 & & 2 & 138 (46.0\%) & 130 (43.3\%) & 32 (10.7\%) & 300 \\
 & & 3 & 103 (34.3\%) & 151 (50.3\%) & 46 (15.3\%) & 300 \\
\midrule
\multirow{12}{*}{11--20} 
 & \multirow{3}{*}{bas-ind} 
   & 1 & 176 (51.8\%) & 164 (48.2\%) & -- & 340 \\
 & & 2 & 130 (38.2\%) & 210 (61.8\%) & -- & 340 \\
 & & 3 & 73 (21.5\%) & 267 (78.5\%) & -- & 340 \\

 & \multirow{3}{*}{bas-net} 
   & 1 & 120 (75.0\%) & 40 (25.0\%) & -- & 160 \\
 & & 2 & 98 (61.3\%) & 62 (38.8\%) & -- & 160 \\
 & & 3 & 42 (26.3\%) & 118 (73.8\%) & -- & 160 \\

 & \multirow{3}{*}{dec-ind} 
   & 1 & 108 (36.0\%) & 175 (58.3\%) & 17 (5.7\%) & 300 \\
 & & 2 & 102 (34.0\%) & 176 (58.7\%) & 22 (7.3\%) & 300 \\
 & & 3 & 54 (18.0\%) & 221 (73.7\%) & 25 (8.3\%) & 300 \\

 & \multirow{3}{*}{dec-net} 
   & 1 & 102 (63.8\%) & 58 (36.2\%) & -- & 160 \\
 & & 2 & 63 (39.4\%) & 83 (51.9\%) & 14 (8.8\%) & 160 \\
 & & 3 & 59 (36.9\%) & 88 (55.0\%) & 13 (8.1\%) & 160 \\
\midrule
\multirow{12}{*}{All Rounds} 
 & \multirow{3}{*}{bas-ind} 
   & 1 & 302 (60.4\%) & 198 (39.6\%) & -- & 500 \\
 & & 2 & 207 (41.4\%) & 293 (58.6\%) & -- & 500 \\
 & & 3 & 131 (26.2\%) & 369 (73.8\%) & -- & 500 \\

 & \multirow{3}{*}{bas-net} 
   & 1 & 341 (68.2\%) & 159 (31.8\%) & -- & 500 \\
 & & 2 & 276 (55.2\%) & 224 (44.8\%) & -- & 500 \\
 & & 3 & 191 (38.2\%) & 309 (61.8\%) & -- & 500 \\

 & \multirow{3}{*}{dec-ind} 
   & 1 & 226 (49.1\%) & 216 (47.0\%) & 18 (3.9\%) & 460 \\
 & & 2 & 166 (36.1\%) & 261 (56.7\%) & 33 (7.2\%) & 460 \\
 & & 3 & 97 (21.1\%) & 332 (72.2\%) & 31 (6.7\%) & 460 \\

 & \multirow{3}{*}{dec-net} 
   & 1 & 270 (58.7\%) & 170 (37.0\%) & 20 (4.3\%) & 460 \\
 & & 2 & 201 (43.7\%) & 213 (46.3\%) & 46 (10.0\%) & 460 \\
 & & 3 & 162 (35.2\%) & 239 (52.0\%) & 59 (12.8\%) & 460 \\
\bottomrule
\end{tabular}
\label{tab:token_decisions_links}
\end{table}

\section{Experimental instructions}\label{instructions}



\section*{Supplementary material}

# B    Experimental instructions

# BASELINE INDIVIDUAL-NETWORK

# Welcome

Thank you for participating in this experiment on decision making.

Please do not communicate with other participants and do not use cell phones or other electronic devices at any point during the experiment. If you have any questions, please raise your hand and an experimenter will come to you to answer your question privately.

This experiment has three parts, Part 1, Part 2 and Part 3.

For your participation in this experiment you will earn a £5 show up fee. You can also earn additional money depending on your decisions in the experiment and other participants' decisions.

During the experiment, all the earnings will be denominated in ECU (Experimental Currency Unit). ECUs will be converted into pounds at the exchange rate 25 ECU = £1. You will receive your payment at the end of the experiment. You will not know the identity of any participant and they will not know yours. No other participant will be informed of your payment and you will not be informed of any other participant's payment.

# Part 1 Instructions

Please give these instructions your full attention. At the end of the instructions, you will be asked some review questions.

**1. Rounds and groups.** Part 1 of the experiment consists of 10 rounds. At the start of the first round, you will be randomly assigned to a group of 6 participants. Your group will remain the same for all the rounds.

**2. Network.** In each round, you and the other five members of your group will be assigned to different positions in a 6-person network. Positions will remain the same for all the rounds. The network is depicted in the figure below with the 6 possible positions labelled A, B, C, D, E and F. In the network, each line represents a link, i.e., a connection between two participants. Note that there are two positions with 1 link (positions A and F), two positions with 2 links (positions C and E) and two positions with 3 links (positions B and D). We will refer to any participant you are linked to as your **neighbour**.

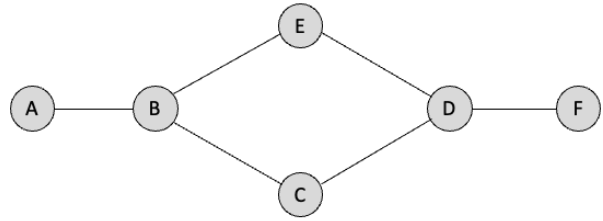

**3. Initial situation** In every round, every participant will start with 150 ECU and will face the **risk of losing 100 ECU**.

Here is how this risk is determined.
Every participant has a box with 100 balls, that can be green, red or brown. At the end of each round, one ball is drawn from each participant's box. If the ball drawn is red or brown, the participant **will lose 100 ECU**. If the ball drawn is green, the participant **will not lose anything.** The balls are drawn independently for each participant.

- 🟤 → you lose 100 ECU
- 🔴 → you lose 100 ECU
- 🟢 → you do not lose anything

Initially, the colours of the balls are as follows.

- Every participant has 25 brown balls

- Every participant has 15 red balls for each link they have. A and F have 15 red balls; C and E have 30 red balls; B and D have 45 red balls.

- The number of green balls is always the difference between 100 and the sum of the brown balls and red balls in the box. E.g., if you have 30 red balls and 25 brown balls, then you will have 45 green balls.

The figure below illustrates the **initial** number of brown, red and green balls for each position.

**Recall that the sum of the red and brown balls determines your risk of losing 100 ECU.**

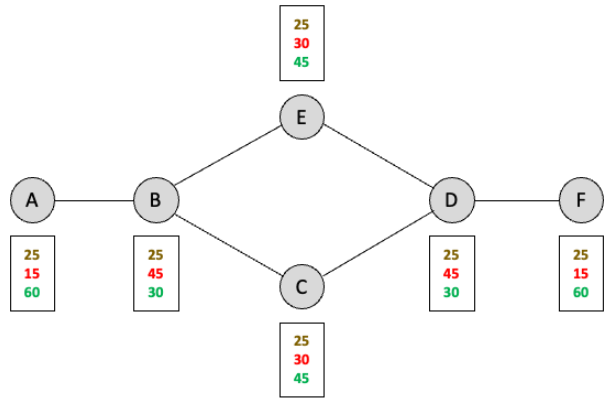

**4. Decision**   In every round, you must decide whether you want to buy token X for the price of 32 ECU or no token. The other members face the same decision as you. **A token reduces your risk of losing 100 ECU**. It reduces the number of red balls in your box by replacing them with green balls.

The following table describes the effects of your possible decisions.

| Your decision | Price | Effect of your decision |
| --- | --- | --- |
| **Buy token X** | 32 ECU | 10 of your own red balls are replaced with 10 green balls for each link you have. |
| **Buy no token** | 0 ECU | No effect |

**The above rules apply to every network member.**

**Your final number of red and brown balls represents your probability of losing 100 ECU.**

After making your decision, you will be asked to guess the other members' decisions. The same applies to the other members.

**5. Round earnings**
If you buy token X for 32 ECU:
- If a red or brown ball is drawn from your box, you lose 100 ECU and your round payoff will be $150 - 32 - 100 = 18$ ECU.
- If a green ball is drawn from your box, your round payoff will be $150 - 32 = 118$ ECU.

3

<u>If you do not buy token X:</u>
- If a red or brown ball is drawn from your box, you lose 100 ECU and your round payoff will be $150 - 100 = 50$.
- If a green ball is drawn from your box, your round payoff will be 150 ECU.

6. **Feedback** In every round, after your decision, you will receive the following information:
- Each participant's decision
- The number of red and brown balls in each participant's box

7. **Payments** At the end of the experiment, 1 out of the 10 rounds in this part of the experiment will be randomly chosen for payment. You will receive the amount that you have earned in this round.

## Part 2 Instructions

Part 2 of the experiment consists of 10 rounds. Your group is the same as in Part 1 and will remain the same for all the 10 rounds in Part 2.

**The only difference with respect to the previous 10 rounds is the following.** Your decision will affect the other members and their decisions will affect you.

The following table describes the effects of your possible decisions.

| Your decision | Price | Effect of your decision |
|---|---|---|
| **Buy token X** | 32 ECU | 10 of your own red balls are replaced with 10 green balls for each link you have, regardless of your neighbours' decisions.<br><br>Each of your neighbours who has not bought token X, will have 10 of their red balls replaced with 10 green balls<br><br>Every member but you will have 5 of their brown balls replaced with 5 green balls. |
| **Buy no token** | 0 ECU | No effect |

**The above rules apply to every network member.** Just like your decisions affect the colour of the balls in other members' boxes, other members' decisions affect the colour of the balls in your box.

**Your final number of red balls depends on your decisions and your neighbours' decisions**, as described in the following table.

4

| Your decision | Effect on your red balls |
|---|---|
| **Buy token X** | 10 of your red balls are replaced with 10 green balls for each link you have, regardless of your neighbours' decisions. |
| **Buy no token** | 10 of your red balls are replaced with 10 green balls for each neighbour who has bought token X.<br><br>No effect if no neighbour buys any token. |

**Your final number of brown balls only depends on the *other* members' decisions.**

- For each member (but you) who buys a token, 5 of your brown balls are replaced with 5 green balls.

**Your final number of red and brown balls represents your probability of losing 100 ECU.**

After making your decision, you will be asked to guess the other members' decisions. The same applies to the other members.

## Part 3

You are now given 40 ECU and must choose the portion of this amount (between 0 ECU and 40 ECU, inclusive) that you wish to invest. The amount not invested is yours to keep. Whatever amount is invested will become 2.5 times larger with 50% chance and will become 0 with 50% chance. Only one participant in this room will be randomly picked and paid for this decision.

How much would you like to invest? \_\_ ECU

\*\*\*\*

In the following table there are 4 rows corresponding to Decision 1, Decision 2, Decision 3 and Decision 4. In each decision you must choose between two options: Option Left (L) and Option Right (R). In each decision the chosen option will specify an earning for you and an earning for another participant in the room. All participants in the room will take the same four decisions.

Once all the participants in the room have taken their decisions, one of the participants will be randomly picked. For the participant that is picked, one of their four decisions will be randomly selected and whatever option they chose in that decision the payments will be made.

Thus, the selected participant will receive the amount of ECU that, according to their option choice in the selected decision, corresponds to them. Additionally, another participant in the room

5

will be randomly selected and will receive the amount of ECU that corresponds to "the other participant".

Please select an option (Option L or Option R) for each decision:

|  | Option L | Option R |
|---|---|---|
| **Decision 1** | ☐ 40 ECU for you / 40 ECU for the other participant | ☐ 40 ECU for you / 20 ECU for the other participant |
| **Decision 2** | ☐ 40 ECU for you / 40 ECU for the other participant | ☐ 60 ECU for you / 20 ECU for the other participant |
| **Decision 3** | ☐ 40 ECU for you / 40 ECU for the other participant | ☐ 40 ECU for you / 80 ECU for the other participant |
| **Decision 4** | ☐ 40 ECU for you / 40 ECU for the other participant | ☐ 60 ECU for you / 100 ECU for the other participant |

\*\*\*\*

Your age: \_\_

Your gender: [female, male, other (non-binary, agender, something else)]

Are you a student? [Yes, No]

Are you employed? [Yes, No]

BASELINE NETWORK-INDIVIDUAL

# Welcome

Thank you for participating in this experiment on decision making.

Please do not communicate with other participants and do not use cell phones or other electronic devices at any point during the experiment. If you have any questions, please raise your hand and an experimenter will come to you to answer your question privately.

This experiment has three parts, Part 1, Part 2 and Part 3.

For your participation in this experiment you will earn a £5 show up fee. You can also earn additional money depending on your decisions in the experiment and other participants' decisions.

During the experiment, all the earnings will be denominated in ECU (Experimental Currency Unit). ECUs will be converted into pounds at the exchange rate 25 ECU = £1. You will receive your payment at the end of the experiment. You will not know the identity of any participant and they will not know yours. No other participant will be informed of your payment and you will not be informed of any other participant's payment.

# Part 1 Instructions

Please give these instructions your full attention. At the end of the instructions, you will be asked some review questions.

**1. Rounds and groups.** Part 1 of the experiment consists of 10 rounds. At the start of the first round, you will be randomly assigned to a group of 6 participants. Your group will remain the same for all the rounds.

**2. Network.** In each round, you and the other five members of your group will be assigned to different positions in a 6-person network. Positions will remain the same for all the rounds. The network is depicted in the figure below with the 6 possible positions labelled A, B, C, D, E and F. In the network, each line represents a link, i.e., a connection between two participants. Note that there are two positions with 1 link (positions A and F), two positions with 2 links (positions C and E) and two positions with 3 links (positions B and D). We will refer to any participant you are linked to as your **neighbour**.

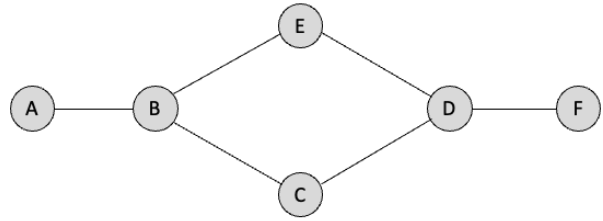

**3. Initial situation** In every round, every participant will start with 150 ECU and will face the **risk of losing 100 ECU**.

Here is how this risk is determined.
Every participant has a box with 100 balls, that can be green, red or brown. At the end of each round, one ball is drawn from each participant's box. If the ball drawn is red or brown, the participant **will lose 100 ECU**. If the ball drawn is green, the participant **will not lose anything.** The balls are drawn independently for each participant.

- 🟤 → you lose 100 ECU
- 🔴 → you lose 100 ECU
- 🟢 → you do not lose anything

Initially, the colours of the balls are as follows.

- Every participant has 25 brown balls

- Every participant has 15 red balls for each link they have. A and F have 15 red balls; C and E have 30 red balls; B and D have 45 red balls.

- The number of green balls is always the difference between 100 and the sum of the brown balls and red balls in the box. E.g., if you have 30 red balls and 25 brown balls, then you will have 45 green balls.

The figure below illustrates the **initial** number of brown, red and green balls for each position.

**Recall that the sum of the red and brown balls determines your risk of losing 100 ECU.**

2

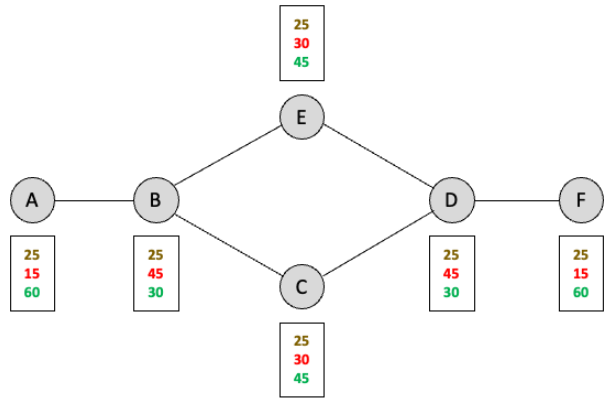

**4. Decision** In every round, you must decide whether you want to buy token X for the price of 32 ECU or no token. The other members face the same decision as you. **A token reduces your risk and other members' risk of losing 100 ECU**. It reduces the number of red balls in your box and in your neighbours' boxes by replacing them with green balls. It also reduces the number of brown balls in the other members' boxes by replacing them with green balls.

The following table describes the effects of your possible decisions.

| Your decision | Price | Effect of your decision |
|---|---|---|
| **Buy token X** | 32 ECU | 10 of your own red balls are replaced with 10 green balls for each link you have, regardless of your neighbours' decisions. Each of your neighbours who has not bought token X, will have 10 of their red balls replaced with 10 green balls. Every member but you will have 5 of their brown balls replaced with 5 green balls. |
| **Buy no token** | 0 ECU | No effect |

**The above rules apply to every network member.** Just like your decisions affect the colour of the balls in other members' boxes, other members' decisions affect the colour of the balls in your box.

**Your final number of red balls depends on your decisions and your neighbours' decisions**, as described in the following table.

3

| Your decision | Effect on your red balls |
|---|---|
| **Buy token X** | 10 of your red balls are replaced with 10 green balls for each link you have, regardless of your neighbours' decisions. |
| **Buy no token** | 10 of your red balls are replaced with 10 green balls for each neighbour who has bought token X.<br><br>No effect if no neighbour buys any token. |

**Your final number of brown balls only depends on the *other* members' decisions.**

- For each member (but you) who buys a token, 5 of your brown balls are replaced with 5 green balls.

**Your final number of red and brown balls represents your probability of losing 100 ECU.**

After making your decision, you will be asked to guess the other members' decisions. The same applies to the other members.

**5. Round earnings**

<u>If you buy token X for 32 ECU:</u>
- If a red or brown ball is drawn from your box, you lose 100 ECU and your round payoff will be $150 - 32 - 100 = 18$ ECU.
- If a green ball is drawn from your box, your round payoff will be $150 - 32 = 118$ ECU.

<u>If you do not buy token X:</u>
- If a red or brown ball is drawn from your box, you lose 100 ECU and your round payoff will be $150 - 100 = 50$.
- If a green ball is drawn from your box, your round payoff will be 150 ECU.

**6. Feedback**   In every round, after your decision, you will receive the following information:

- Each participant's decision
- The number of red and brown balls in each participant's box

**7. Payments**   At the end of the experiment, 1 out of the 10 rounds in this part of the experiment will be randomly chosen for payment. You will receive the amount that you have earned in this round.

## Part 2 Instructions

Part 2 of the experiment consists of 10 rounds. Your group is the same as in Part 1 and will remain the same for all the 10 rounds in Part 2.

**The only difference with respect to the previous 10 rounds is the following.** Your decision will not affect the other members and their decisions will not affect you.

The following table describes the effects of your possible decisions.

| Your decision | Price | Effect of your decision |
|---|---|---|
| **Buy token X** | 32 ECU | 10 of your own red balls are replaced with 10 green balls for each link you have. |
| **Buy no token** | 0 ECU | No effect |

**The above rules apply to every network member.**

**Your final number of red and brown balls represents your probability of losing 100 ECU.**

After making your decision, you will be asked to guess the other members' decisions. The same applies to the other members.

## Part 3

*Same as for the other treatments.*

5

DECOY INDIVIDUAL-NETWORK

# Welcome

Thank you for participating in this experiment on decision making.

Please do not communicate with other participants and do not use cell phones or other electronic devices at any point during the experiment. If you have any questions, please raise your hand and an experimenter will come to you to answer your question privately.

This experiment has three parts, Part 1, Part 2 and Part 3.

For your participation in this experiment you will earn a £5 show up fee. You can also earn additional money depending on your decisions in the experiment and other participants' decisions.

During the experiment, all the earnings will be denominated in ECU (Experimental Currency Unit). ECUs will be converted into pounds at the exchange rate 25 ECU = £1. You will receive your payment at the end of the experiment. You will not know the identity of any participant and they will not know yours. No other participant will be informed of your payment and you will not be informed of any other participant's payment.

# Part 1 Instructions

Please give these instructions your full attention. At the end of the instructions, you will be asked some review questions.

**1. Rounds and groups.** Part 1 of the experiment consists of 10 rounds. At the start of the first round, you will be randomly assigned to a group of 6 participants. Your group will remain the same for all the rounds.

**2. Network.** In each round, you and the other five members of your group will be assigned to different positions in a 6-person network. Positions will remain the same for all the rounds. The network is depicted in the figure below with the 6 possible positions labelled A, B, C, D, E and F. In the network, each line represents a link, i.e., a connection between two participants. Note that there are two positions with 1 link (positions A and F), two positions with 2 links (positions C and E) and two positions with 3 links (positions B and D). We will refer to any participant you are linked to as your **neighbour**.

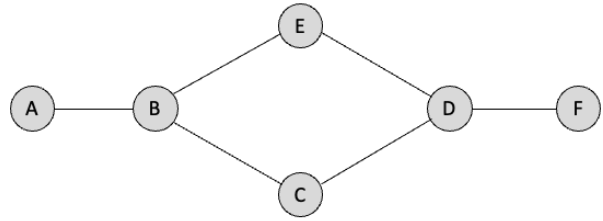

## 3. Initial situation
In every round, every participant will start with 150 ECU and will face the **risk of losing 100 ECU**.

Here is how this risk is determined.
Every participant has a box with 100 balls, that can be green, red or brown. At the end of each round, one ball is drawn from each participant's box. If the ball drawn is red or brown, the participant **will lose 100 ECU**. If the ball drawn is green, the participant **will not lose anything.** The balls are drawn independently for each participant.

- 🟤 → you lose 100 ECU
- 🔴 → you lose 100 ECU
- 🟢 → you do not lose anything

Initially, the colours of the balls are as follows.

- Every participant has 25 brown balls

- Every participant has 15 red balls for each link they have. A and F have 15 red balls; C and E have 30 red balls; B and D have 45 red balls.

- The number of green balls is always the difference between 100 and the sum of the brown balls and red balls in the box. E.g., if you have 30 red balls and 25 brown balls, then you will have 45 green balls.

The figure below illustrates the **initial** number of brown, red and green balls for each position.

**Recall that the sum of the red and brown balls determines your risk of losing 100 ECU.**

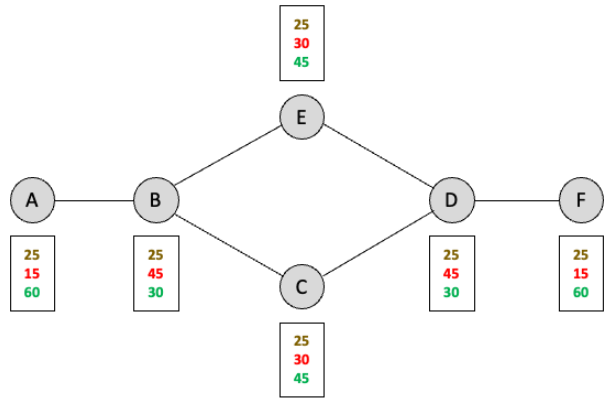

**4. Decision** In every round, you must decide whether you want to buy token X for the price of 32 ECU, token Y for the price of 42 ECU, or no token. The other members face the same decision as you. **A token reduces your risk of losing 100 ECU**. It reduces the number of red balls in your box by replacing them with green balls.

The following table describes the effects of your possible decisions.

| Your decision | Price | Effect of your decision |
|---|---|---|
| **Buy token X** | 32 ECU | 10 of your red balls are replaced with 10 green balls for each link you have. |
| **Buy token Y** | 42 ECU | 8 of your red balls are replaced with 8 green balls for each link you have, |
| **Buy no token** | 0 ECU | No effect |

**The above rules apply to every network member.**

**Your final number of red and brown balls represents your probability of losing 100 ECU.**

After making your decision, you will be asked to guess the other members' decisions. The same applies to the other members.

**5. Round earnings**
If you buy token X for 32 ECU:

3

- If a red or brown ball is drawn from your box, you lose 100 ECU and your round payoff will be $150 - 32 - 100 = 18$ ECU.
- If a green ball is drawn from your box, your round payoff will be $150 - 32 = 118$ ECU.

If you buy token Y for 42 ECU:
- If a red or brown ball is drawn from your box, you lose 100 ECU and your round payoff will be $150 - 42 - 100 = 9$ ECU.
- If a green ball is drawn from your box, your round payoff will be $150 - 42 = 109$ ECU.

If you do not buy either token:
- If a red or brown ball is drawn from your box, you lose100 ECU and your round payoff will be $150 - 100 = 50$.
- If a green ball is drawn from your box, your round payoff will be 150 ECU.

6. **Feedback** In every round, after your decision, you will receive the following information:
- Each participant's decision
- The number of red and brown balls in each participant's box

7. **Payments** At the end of the experiment, 1 out of the 10 rounds in this part of the experiment will be randomly chosen for payment. You will receive the amount that you have earned in this round.

## Part 2 Instructions

Part 2 of the experiment consists of 10 rounds. Your group is the same as in Part 1 and will remain the same for all the 10 rounds in Part 2.

<span style="color:red">The only difference with respect to the previous 10 rounds is the following.</span>

The following table describes the effects of your possible decisions.

**The above rules apply to every network member.** Just like your decisions affect the colour of the balls in other members' boxes, other members' decisions affect the colour of the balls in your box.

**Your final number of red balls depends on your decisions and your neighbours' decisions**, as described in the following table.

**Your final number of brown balls only depends on the *other* members' decisions.**

4

| Your decision | Price | Effect of your decision |
|---|---|---|
| **Buy token X** | 32 ECU | 10 of your red balls are replaced with 10 green balls for each link you have, regardless of your neighbours' decisions. <br><br> Each of your neighbours who has not bought either token X or Y will have 10 of their red balls replaced with 10 green balls. <br><br> Each of your neighbours who has bought token Y, will have 2 of their red balls replaced with 2 green balls. <br><br> Every member but you will have 5 of their brown balls replaced with 5 green balls. |
| **Buy token Y** | 42 ECU | 8 of your red balls are replaced with 8 green balls for each link you have, regardless of your neighbours' decision. <br><br> Each of your neighbours who has not bought either token X or Y will have 8 of their red balls replaced with 8 green balls. <br><br> Every member but you will have 5 of their brown balls replaced with 5 green balls. |
| **Buy no token** | 0 ECU | No effect |

| Your decision | Effect on your red balls |
|---|---|
| **Buy token X** | 10 of your red balls are replaced with 10 green balls for each link you have, regardless of your neighbours' decisions. |
| **Buy token Y** | 10 of your red balls are replaced with 10 green balls for each neighbour who has bought token X. <br><br> 8 of your red balls are replaced with 8 green balls for each neighbour who has bought token Y or no token. |
| **Buy no token** | 10 of your red balls are replaced with 10 green balls for each neighbour who has bought token X. <br><br> 8 of your red balls are replaced with 8 green balls for each neighbour who has bought token Y. <br><br> No effect if no neighbour buys any token. |

- For each member (but you) who buys a token, 5 of your brown balls are replaced with 5 green balls.

**Your final number of red and brown balls represents your probability of losing 100 ECU.**

After making your decision, you will be asked to guess the other members' decisions. The same

applies to the other members.

# Part 3

*Same as for the other treatments*

DECOY NETWORK-INDIVIDUAL

# Welcome

Thank you for participating in this experiment on decision making.

Please do not communicate with other participants and do not use cell phones or other electronic devices at any point during the experiment. If you have any questions, please raise your hand and an experimenter will come to you to answer your question privately.

This experiment has three parts, Part 1, Part 2 and Part 3.

For your participation in this experiment you will earn a £5 show up fee. You can also earn additional money depending on your decisions in the experiment and other participants' decisions.

During the experiment, all the earnings will be denominated in ECU (Experimental Currency Unit). ECUs will be converted into pounds at the exchange rate 25 ECU = £1. You will receive your payment at the end of the experiment. You will not know the identity of any participant and they will not know yours. No other participant will be informed of your payment and you will not be informed of any other participant's payment.

# Part 1 Instructions

Please give these instructions your full attention. At the end of the instructions, you will be asked some review questions.

**1. Rounds and groups.** Part 1 of the experiment consists of 10 rounds. At the start of the first round, you will be randomly assigned to a group of 6 participants. Your group will remain the same for all the rounds.

**2. Network.** In each round, you and the other five members of your group will be assigned to different positions in a 6-person network. Positions will remain the same for all the rounds. The network is depicted in the figure below with the 6 possible positions labelled A, B, C, D, E and F. In the network, each line represents a link, i.e., a connection between two participants. Note that there are two positions with 1 link (positions A and F), two positions with 2 links (positions C and E) and two positions with 3 links (positions B and D). We will refer to any participant you are linked to as your **neighbour**.

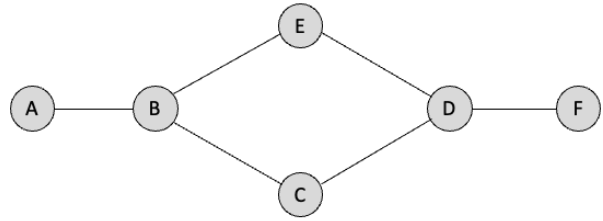

**3. Initial situation** In every round, every participant will start with 150 ECU and will face the **risk of losing 100 ECU**.

Here is how this risk is determined.
Every participant has a box with 100 balls, that can be green, red or brown. At the end of each round, one ball is drawn from each participant's box. If the ball drawn is red or brown, the participant **will lose 100 ECU**. If the ball drawn is green, the participant **will not lose anything.** The balls are drawn independently for each participant.

- 🟤 → you lose 100 ECU
- 🔴 → you lose 100 ECU
- 🟢 → you do not lose anything

Initially, the colours of the balls are as follows.

- Every participant has 25 brown balls

- Every participant has 15 red balls for each link they have. A and F have 15 red balls; C and E have 30 red balls; B and D have 45 red balls.

- The number of green balls is always the difference between 100 and the sum of the brown balls and red balls in the box. E.g., if you have 30 red balls and 25 brown balls, then you will have 45 green balls.

The figure below illustrates the **initial** number of brown, red and green balls for each position.

**Recall that the sum of the red and brown balls determines your risk of losing 100 ECU.**

2

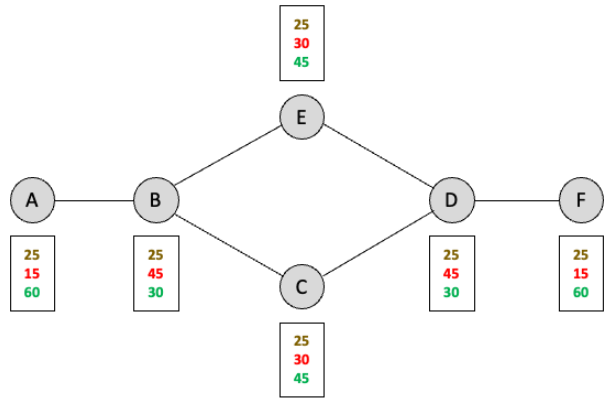

**4. Decision** In every round, you must decide whether you want to buy token X for the price of 32 ECU, token Y for the price of 42 ECU, or no token. The other members face the same decision as you. **A token reduces your risk and other members' risk of losing 100 ECU**. It reduces the number of red balls in your box and in your neighbours' boxes by replacing them with green balls. It also reduces the number of brown balls in the other members' boxes by replacing them with green balls.

The following table describes the effects of your possible decisions.

| Your decision | Price | Effect of your decision |
|---|---|---|
| **Buy token X** | 32 ECU | 10 of your red balls are replaced with 10 green balls for each link you have, regardless of your neighbours' decisions.<br><br>Each of your neighbours who has not bought either token X or Y will have 10 of their red balls replaced with 10 green balls.<br><br>Each of your neighbours who has bought token Y, will have 2 of their red balls replaced with 2 green balls.<br><br>Every member but you will have 5 of their brown balls replaced with 5 green balls. |
| **Buy token Y** | 42 ECU | 8 of your red balls are replaced with 8 green balls for each link you have, regardless of your neighbours' decision.<br><br>Each of your neighbours who has not bought either token X or Y will have 8 of their red balls replaced with 8 green balls.<br><br>Every member but you will have 5 of their brown balls replaced with 5 green balls. |
| **Buy no token** | 0 ECU | No effect |

**The above rules apply to every network member.** Just like your decisions affect the colour

3

of the balls in other members' boxes, other members' decisions affect the colour of the balls in your box.

**Your final number of red balls depends on your decisions and your neighbours' decisions**, as described in the following table.

| Your decision | Effect on your red balls |
|---|---|
| **Buy token X** | 10 of your red balls are replaced with 10 green balls for each link you have, regardless of your neighbours' decisions. |
| **Buy token Y** | 10 of your red balls are replaced with 10 green balls for each neighbour who has bought token X.<br><br>8 of your red balls are replaced with 8 green balls for each neighbour who has bought token Y or no token. |
| **Buy no token** | 10 of your red balls are replaced with 10 green balls for each neighbour who has bought token X.<br><br>8 of your red balls are replaced with 8 green balls for each neighbour who has bought token Y.<br><br>No effect if no neighbour buys any token. |

**Your final number of brown balls only depends on the *other* members' decisions.**

- For each member (but you) who buys a token, 5 of your brown balls are replaced with 5 green balls.

**Your final number of red and brown balls represents your probability of losing 100 ECU.**

After making your decision, you will be asked to guess the other members' decisions. The same applies to the other members.

**5. Round earnings**

<u>If you buy token X for 32 ECU:</u>
- If a red or brown ball is drawn from your box, you lose 100 ECU and your round payoff will be $150 - 32 - 100 = 18$ ECU.
- If a green ball is drawn from your box, your round payoff will be $150 - 32 = 118$ ECU.

<u>If you buy token Y for 42 ECU:</u>

- If a red or brown ball is drawn from your box, you lose 100 ECU and your round payoff will be $150 - 42 - 100 = 9$ ECU.
- If a green ball is drawn from your box, your round payoff will be $150 - 42 = 109$ ECU.

If you do not buy either token:
- If a red or brown ball is drawn from your box, you lose 100 ECU and your round payoff will be $150 - 100 = 50$.
- If a green ball is drawn from your box, your round payoff will be 150 ECU.

6. **Feedback** In every round, after your decision, you will receive the following information:
- Each participant's decision
- The number of red and brown balls in each participant's box

7. **Payments** At the end of the experiment, 1 out of the 10 rounds in this part of the experiment will be randomly chosen for payment. You will receive the amount that you have earned in this round.

## Part 2 Instructions

Part 2 of the experiment consists of 10 rounds. Your group is the same as in Part 1 and will remain the same for all the 10 rounds in Part 2.

**The only difference with respect to the previous 10 rounds is the following.** Your decision will not affect the other members and their decisions will not affect you.

The following table describes the effects of your possible decisions.

| Your decision | Price | Effect of your decision |
|---|---|---|
| **Buy token X** | 32 ECU | 10 of your red balls are replaced with 10 green balls for each link you have |
| **Buy token Y** | 42 ECU | 8 of your red balls are replaced with 8 green balls for each link you have. |
| **Buy no token** | 0 ECU | No effect |

**The above rules apply to every network member.**

**Your final number of red and brown balls represents your probability of losing 100 ECU.**

After making your decision, you will be asked to guess the other members' decisions. The same applies to the other members.

# Part 3

*Same as for the other treatments*



\end{document}